# FREQUENCY DISTRIBUTION OF ACOUSTIC OSCILLATION IN THE SOLAR ATMOSPHERE DURING FLARE EVENT

A. Wiśniewska[1,3] E. Chmielewska[2], K. Radziszewski[2], M. Roth[3], J. Staiger[3]

1. Physics Institute, Jagiellonian University,
ul. Lojasiewicza 11, 30-348 Krakow, Poland
2. Astronomical Institute, University of Wroclaw,
ul. Kopernika 11, 51-622 Wroclaw, Poland
3. Leibniz-Institut für Sonnenphysik (KIS),
Schöneckstr. 6, D-79 104 Freiburg, Germany

e-mail: awisniewska@aip.de

**Abstract:**
We present a study of multi-wavelength observations, of a **C 2.3** solar flare in Active Region **NOAA 12353, observed on 2015 May 23**, which reveal new properties of acoustic waves in the flaring region. The space-, and ground-based data measured by the HELioseismological Large Regions Interferometric Device, operating at the Vacuum Tower Telescope, the Atmospheric Imaging Assembly, and Helioseismic and Magnetic Imager on board the Solar Dynamic Observatory, were used in this paper.

First, using power spectra of solar oscillations, we identified the dominant frequencies and their location at seven different atmospheric levels before and after the flare event. Second, based on AIA observations taken in six Extreme Ultraviolet filters, we derived Differential Emission Measure (DEM) profiles and DEM maps of the flare. Finally, we confirm the σ-classification of the magnetic field in the active area, directly related to the flare. Our results are as follows: the high-frequency waves ($\nu > 5$ mHz) in the photosphere, in both cases, before and after the flare, are generated at the foot-points of the chromospheric loop, while in the chromosphere (H$\alpha$ line), before the event the power enhancement exhibits for maximum of flare emission, and after the eruption the enhancement by all frequencies is observed only in the post flare loop area. Moreover, the power of oscillation in the pores surrounding area before the flare has a random character, while after the flare oscillation's power is concentrated in the pore, and weakened outside of.

We conclude that the accurate detection of high-frequency acoustic waves in the active regions can lead to faster and easier prediction of high-energy events.

## 1. INTRODUCTION

Despite of many investigations of the flares in the solar corona and the middle chromosphere, their effect on the background solar has oscillations still not been well explained.

Intense solar flares manifest itself at all wavelengths of electromagnetic radiation, from radio waves up to gamma-rays range. Sudden release of the magnetic field energy in the solar corona entails many explosive disturbances, which often also occur in the chromosphere and sometimes even in the photosphere (Priest (1982); Aschwanden (2005)). A quick increase, in the time range of seconds, of plasma temperature and pressure, as well as chromosphere heating by downward particle beams moving with relativistic velocities along magnetic structures, causes the solar flares. Flares are the most energetic and disturbing phenomena observed on the Sun. The msulti-temperature plasma distribution, propagation of the shock waves, and the sub-second time-scale of physical processes occurring during flares require multi-wavelength observations with high spatial and time resolutions are essential for the solar flare research.

During solar flares, large number of non-thermal electrons accelerate in the solar corona, and on timescale of seconds reach the dense chromosphere. Accelerated particles could transport 10% - 50% of the total energy released in corona during a flare (Lin et al. (2002)). These particles deposit their energy in the chromosphere, which produces bremsstrahlung hard X-ray emission, and heats the chromospheric plasma (Korchak (1967); Brown (1971, 1972); Hudson (1972)). The response of the heated chromosphere is very rapid, as seen in a range of chromospheric lines like H$\alpha$ line (Kurokawa et al. (1986); Trottet et al. (2000); Wang et al. (2000); Radziszewki et al. (2011)). Additionally, thermal conduction can enough energy to heat the chromosphere on a longer time scale than short-lasting particle beams. That is why a large portion of the flare emission originates from the solar chromosphere. Initially, the flare takes place in the solar corona. Reconnection of the magnetic field in the high corona region leads to the creation of loops filled with hot plasma. However, the coronal density of plasma is too low to explanation the observed electron densities of flaring loops on the level of $10^{10}$ - $10^{11}$ cm$^{-3}$. After sudden heating of the foot-points of loops (by electron beams), chromospheric evaporation fills flaring loops with additional plasma, increasing their densities. Evaporating plasma can be extra heated by gas-dynamic shocks and slow MHD shocks (Cargill and Priest (1983)). Consecutive magnetic field reconnection can create a system of flaring loops, in which the hottest loop is recent and cooler ones are older - due to energy losses by electromagnetic emission (Kopp & Pneuman (1976)). Hot loops filled by plasma with temperatures from a few up to 30 MK are well visible in soft X-ray emission, while most cooling loops are well visible in the extreme ultraviolet EUV range (Benz, (2017)). The coldest loops (temperature below 12 kK) begin to be visible in chromospheric lines (like Hα) and are known as a post- flare loops (Tandberg-Hansen (1995); Vial (1998)). During the decay phase of the flare, cooled plasma flows downward along the post- flare loops toward the chromosphere and can sometimes be lasting even a few hours. In this work, we will be investigating flare event at many different atmospheric levels.

Simultaneous multi-height observations allow us identify coupling between atmospheric layers. The Helioseismological Large Region Interferometric Device (HELLRIDE) instrument, mounted at the Vacuum Tower Telescope (VTT) at the Observatorio del Teide, as one of the multi-wavelength spectrometers, is dedicated to such observations (see Staiger (2012)). In previous publications we reported already about the properties of solar oscillations in the quiet-Sun region and estimated the height dependence of the acoustic cutoff frequency in the non-magnetized (B < 50 G) atmosphere (Wisniewska et al. (2016); Murawski et al. (2016)). During solar flares large number of energetic electrons are accelerated in the solar corona, releasing their energy in the chromosphere.
A large number of observations were taken employing the Hα line (Harvey et al. (1993), Jain et al. (1999)). The results from these observations confirm the existence of chromospheric oscillations with the frequencies of 3.5 mHz and 5.6 mHz excited by flares. Beyond that, solar flares can are visible in many different spectral lines. Despite the fact that this event occurs in the chromospheric line range, several authors claim that the origin of the flare phenomenon is not only magnetic reconnection, taking place in the solar corona, but it might occur as a result of changes in the magnetic field in low atmospheric layers, which implicates solar photosphere as well (Kosovichev & Zharkova (2001); Gossain (2012)). For instance, Sudol & Harvey (2005) investigated a 15-ten X-class solar flare. In 75% of these events, the photospheric magnetic field change occurred in time less than 10 minutes. It is also known, that in very energetic events classified as X-class flares, the particle emission manifests itself in the photosphere and the lower chromosphere in the form of the white-light flares (Boyer et al. (1985); Chen & Ding (2006)). Based on that effect, the focus of our study is the oscillations generated in the photosphere and chromosphere, which propagate through the atmospheric layers due to the flare phenomenon.

Helioseismic measurements in the past have proven that the quiet solar photosphere is dominated by 5 minute oscillations (Leighton (1962); Deubner (1970); Claverie (1981); Deubner et al. (1990, 1992)), whose power maximum lies in the frequency range of 3-4 mHz. However, the flare event occurs only in active regions, very often in well-developed sunspots, for which the dominant frequency is so far not estimated. As mentioned for instance, by Khomenko & Collados (2006), in 'sunspot-like' regions the slow (acoustic) mode continues propagating upward to the chromosphere along the magnetic field lines with increasing amplitude. Also, as reported by Braun et al. (1987) in the sunspots, on a photospheric level, the acoustic power of outgoing waves is strongly reduced when compared with the incoming oscillations. Such behavior is characteristic for a wide range of wave frequencies. Therefore, it is expected to observe the propagation of high-frequency oscillations in the upper levels of the photosphere. In the case of an eruptive flare event, because of its origin, the mode conversion and distribution differ from the normal sunspot oscillations. This problem becomes currently prominent in helioseismic analysis. The work of Monsue et al. (2016), discusses oscillations for M- and X-class flares, as measured by the Global Oscillation Network Group instrument. The authors investigated solar waves with frequencies between 0 and 8 mHz using the Hα line. Their results show power enhancement in time for frequencies of 1-2 mHz immediately in prior and shortly after the flare. The other frequencies, up to the 8 mHz were suppressed. Their explanation for such behavior of the waves was that acoustic energy is being converted into thermal energy at flare maximum, while the low-frequency enhancement may arose from instability in the chromosphere and provided an early warning of the flare onset. The analysis in Monsue et al. (2016) is based mainly on differences between intensity- and velocity oscillations (I-V cross-spectra).

In our work, we widen the scope of investigations, using multi-channel observations. We trace acoustic oscillations during a flare from the bottom photosphere to the chromosphere. The averaged height spacing between the spectral lines is about 200 km. We focus on answering the question of the behavior of high-frequency waves (5 mHz<ν <14 mHz) propagation along the magnetic field lines before, during and in the gradual phase of the flare. For this purpose, we used a 3D-Fourier transform. In this work, we are presenting results of observations taken in the active region NOAA12353, on 23 of May 2015 with the occurring flare event. This paper is organized as follows: In Section 2, we describe the observational data and methods of analysis. We discuss the evolution of the flare phenomenon. The frequency distribution of solar oscillations before and after the event is presented in Section 3. The reconstruction of the magnetic field in the active region is performed in Section 4, and our conclusions are given in Section 5.

## 2. OBSERVATIONS AND DATA ANALYSIS

### 2.1. Multi-wavelength observations Taken with the HELLRIDE

The multi-wavelength observations taken with the HELLRIDE allows probing both the photospheric layer from 200 km (represented by several iron lines) and the very high chromospheric level of around 2000 km in Hα, with very accurate probing of 28 sec. Measurements are fundamental for the construction of the 3D picture of the active region, and provide the information about the plasma behavior at many different atmospheric levels, and in each point of the observed area.

The flare occurred on 2015 May 23 in the active region NOAA 12353 (with two evolving pores). The total length of spectroscopic measurement was 4.15 hr. The data contains 3 hr of a pre-flare stage in this region and the next 1.15 hr with an eruptive flare event. Observations were corrected by the KIS Adaptive Optic System (KAOS) (see von der Luehe et al. 2003). This active region was close to the disk center that day. The position at the beginning of the measurement, in heliographic coordinates at 14:26:00 was ( x:168", y:148" ).

The measurements were taken in seven spectral ranges, each with the cadence of 28 sec. The high temporal cadence of the measurement allows studying oscillations with frequencies up to 17 mHz. Typical solar atmospheric oscillations have a frequency maximum of 3.5 mHz for the photospheric levels. In the chromosphere for the waves, this maximum can also be achieved at the frequencies even up to 7 mHz (Wisniewska et al. (2016)). In the case of the observed active region, which contains two small pores with complex magnetic field structures, exact estimation of the atmospheric formation heights for each spectral line is very hard. However, in the cores of some spectral lines, we observed the presence of specific features like plagues (e.g. Mg I $b_2$), and fibrils (e.g. Hα), which are characteristic at given heights, therefore we concluded that the order of atmospheric heights could be comparable with the results calculated for quiet-Sun regions presented in Wisniewska et al. (2016). Therefore, we took observations in four low photospheric lines Ni I 5435 Å, Fe I 6173 Å, Fe I 6302 Å, at heights of around 230 km, 320 km, and 350 km respectively. The Mg I $b_2$ at 5173 Å and Fe I 5434 Å lines formed around 600 km above the solar surface, representing the middle photosphere. The lines Na $D_2$ at 5890 Å and H α 6562 Å are mostly using for probing the chromospheric spectral regime from 900 up to 2000 km ( cf. Table 1 in Wisniewska et al., (2016)). For each of those measurements, the dopplergram was calculated for each pixel of the image, using the center-of-gravity algorithm (Rees \& Semel, (1979)). The dopplergrams contain the Line-of-Sight velocities of the plasma $V_{LoS}$. At measured velocities, we employed the 3D-Fourier transform, to obtain a density function $F_{Vi}$, which provides the information about the energy per unit frequency. Spectral analysis the follows equations described in Wisniewska et al. (2016). To obtain a periodogram or spectral energy density at a given frequency, we calculated the square modulus of the Fourier transform as:

$$P_{Vi} = |F_{Vi}|^2 , \qquad (1)$$

where *i* represents each measured wavelength. The periodogram is also often called a 'power spectrum'.

For each spectral line measurement, we selected chosen frequency bands in intervals of 3-4, 5-6, 6-7, 8-9, 10-12 and 12-14 mHz. This selection shows the power distribution of oscillations at given frequency ranges. After implementation of an inverse Fourier transform for the data sets, we obtain 2D power map, which allows us to identify dominant frequencies at given atmospheric height in the measured area. Moreover, we can precisely discriminate at which point of the determined region propagates waves in a given frequency band. Furthermore, it is possible to observe which wave frequencies are strongest compared to the other maps with different frequency ranges.

*2.2. AIA Observations and DEM analysis*

In this study, we used Solar Dynamic Observatory (SDO)/AIA images with high spatial (0.6 arcsec pixel$^{-1}$) and temporal (12 sec) resolutions (Lemen et al. 2012). The broad thermal range of AIA observations (from 0.6 to 32 MK) enables reconstruction of DEM profiles for the flare event (O'Dwyer et al. 2010). To perform the DEM analysis, the following six of the AIA filters were used: 94, 131, 171, 193, 21, and 335 Å. For those EUV bandpass, we derived AIA response functions using the CHIANTI database in a broad temperature range (log(T): 5.5 -7.5). The data relevant for the flare event, before processing by AIA_PREP procedure, were deconvolved with the point-spread function in each filter using AIA_DECONVOLVE\_RICHARDSONLUCY. Based on the SDO/Helioseismic and Magnetic Imager (HMI) (Schou et al. 2012 a, 2012 b) LOS magnetograms with a time cadence of 45 sec and spatial resolution 1.0 arcsec we reconstructed magnetic field lines for the active region in the gradual phase of the flare. The DEM is a physical quantity that informs about the plasma distribution with temperature along the LOS *h*. Recovering DEM profiles cannot be determined directly, in a simple way from the multi-wavelength observations. The observed intensities for each filter, $I_n$, can be defined as a convolution of the instrumental response function $R_n(T)$ and the DEM ($\xi(T)$), and it is disturbed by random noise ($\delta n_n$):

$$I_n = \int R_n(T)\xi(T)dT + \delta n_n, \qquad (2)$$

There are many methods to determine the DEMs for the EUV observations taken by AIA. In this paper, we calculated DEM profiles for selected box 2x2 pixels and DEM maps for each pixels using two methods (Chmielewska et al. 2016): the enhanced regularized inversion implemented by Hannah & Kontar (2012), and the iterative forward-fitting method (XRT method) developed initially for HINODE/XRT data by Mark Weber (e.g. Weber et al.(2004)). Using those two DEM calculation techniques improved the interpretations of results and understanding the thermal properties of the flare. The XRT method is available in the SolarSoftWare (SSW) and was successfully implemented for use with the AIA filters (see appendix of Cheng et al. (2012)). An iterative forward-fitting method minimizes the differences in the fitted and the observed intensities in the six EUV AIA bandpass. With this technique, the DEM uncertainties are calculated by the 100 Monte Carlo (MC) realizations disturbed by random noise. The second DEM method used only in a recovering the DEM profiles is a regularized inversion technique initially developed for Reuven Ramaty High Energy Solar Spectroscopic Imager (RHESSI) spectra. This technique calculates both emission measure and temperature error bars. Using the RI and the XRT methods, we reconstructed DEM profiles for two distinct regions in the AIA images for the flare event. We also created DEM maps where the DEM is recovered for each pixel in the six channels for the logarithmic temperature ranges from 5.5 up to 7.5. DEM maps are calculated using only the XRT method.

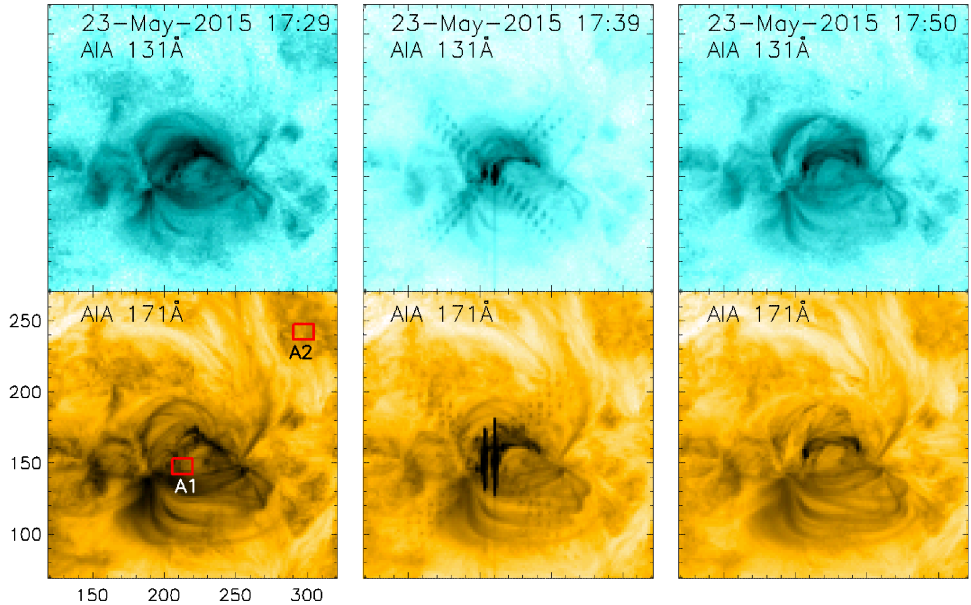

*Figure 1.* AIA 131 Å and 171 Å images of the evolution of the flare that occurred on 2015 May 23, of three moments in time. The red boxes show the selected regions used for the subsequent DEMs analysis.

### 2.3 DEM analysis for the flare event

On 2015 May 23, a C 2.3 class flare was observed in AIA images. Based on the data from the Geostationary Operational Enviromantal Satellite (GOES) the flare event in active region NOAA 12353, on 2015 of May 23 at 17:30:00 UT. The peak of the flare was at 17:39:00 UT and it ended 17:41:00 UT. The evolution of the flare was visible in all the EUV bandpasses (a sequence of the AIA images in ranges of 131 and 171 Å ranges is presented in the Figure 1).

A DEM analysis was carried out for two moments (before the flare onset and during the peak time GOES soft X-ray flux) for the two selected regions defined as follows: (A1) - the 2 by 2 box where the high frequency was recognized and (A2) the 2 by 2 quiet-Sun box. The DEM profiles recovered for A1 structure revealed two distinct components: one was in the temperature range 0,8 - 2,5 MK, the second peaked in 20–25 MK, both in the emission range $10^{19}$-$10^2$ cm$^{-5}$ K$^{-1}$. Those two maxima were seen, especially in the curve obtained by the RI method. The lower temperature seems to be associated with the emission of the background corona along the LOS, whereas the higher temperature is from the flare loop. Before the flare onset, the DEM profile has a weaker low-temperature component and a much weaker high-temperature peak, and the highest temperature range is zero within the error bars (Fig. 2). The observed variances for the RI and the XRT methods seem to be an effect of the different approaches used in those calculation techniques. However, in the case of those two methods, the second high-temperature component is seen during the peak time of the flare as well as DEM profiles for A1 region have higher values. The DEM profiles (Fig. 3) for the A2 region showed a weak emission and narrow DEM profiles in both methods. In the emission range $10^{19}$-$10^{23}$ cm$^5$ K$^{-1}$, this structure ranges from 0.8 up to 2.5 MK. In the DEM profiles before the flare started, a hot component was seen, but it was not included in the further analysis due to the large errors. The pre-flare DEM performed a weaker emission owing to the lower density before the flare onset. Using the XRT method, we prepared DEM maps for the moment in time where the GOES flux reached the maximum. To reduce of random noise of DEM calculations, we reduced spatial resolution by resizing the images by a factor of two. Afterward, we calculated DEM values for each new pixel based on intensities measured in six EUV bands. The resulting maps showed emission in the broad range of temperatures, revealed the complex, multi-thermal nature of the flare (Fig.1 ).

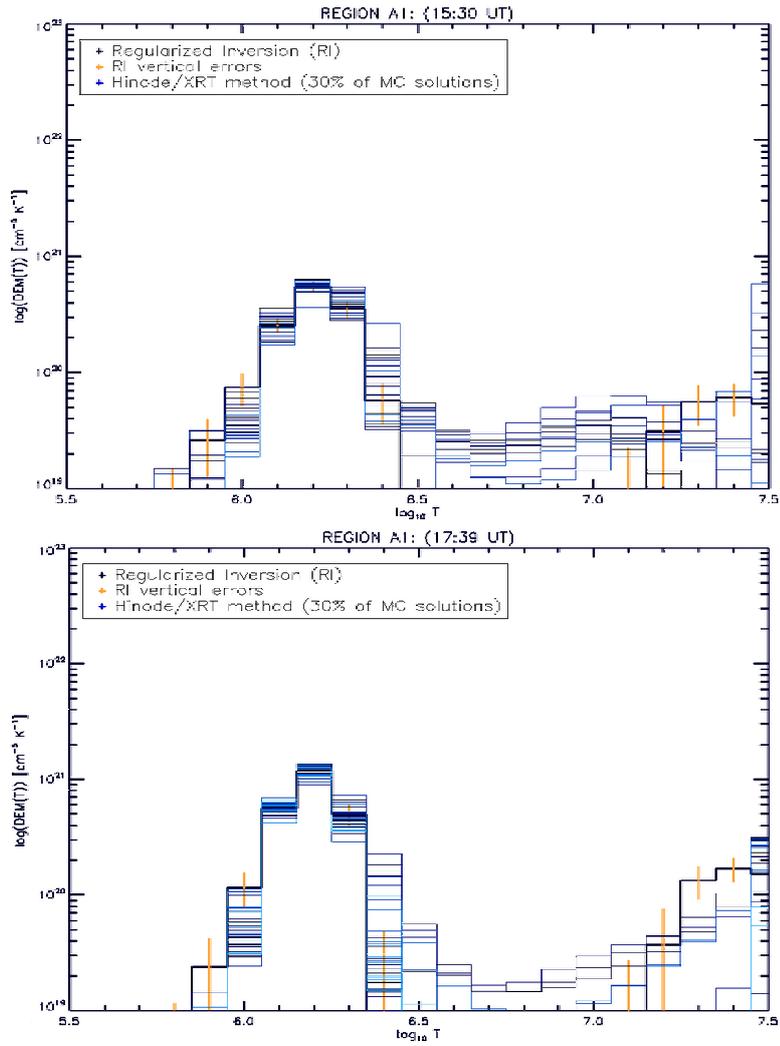

*Figure 2.* DEM profiles of the *A1* region (shown by the red square in Fig. 1 in two moments in time- 15:30 UT - before the flare onset and during the peak of GOES flux. The DEM profiles recovered by RI (bold, black solid line with the red, vertical error bars) and the XRT method (the blue, dashed lines are 20 realizations of MC).

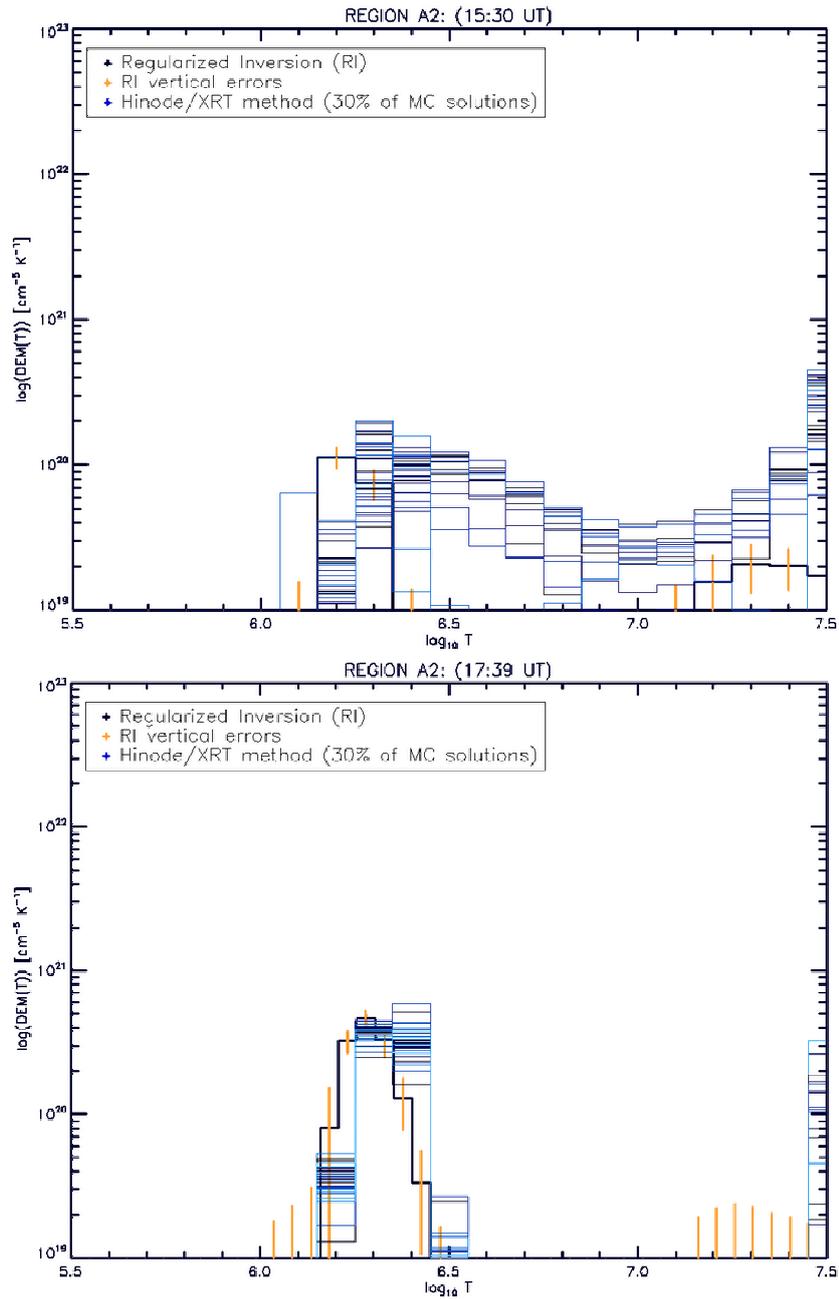

*Figure 3.* DEM profiles of the *A2* region (shown by the red square in Fig. 1 ) in two moments in time- 15:30 UT - before the flare onset and during the peak of GOES flux. The color legend is the same as in the Fig. 2.

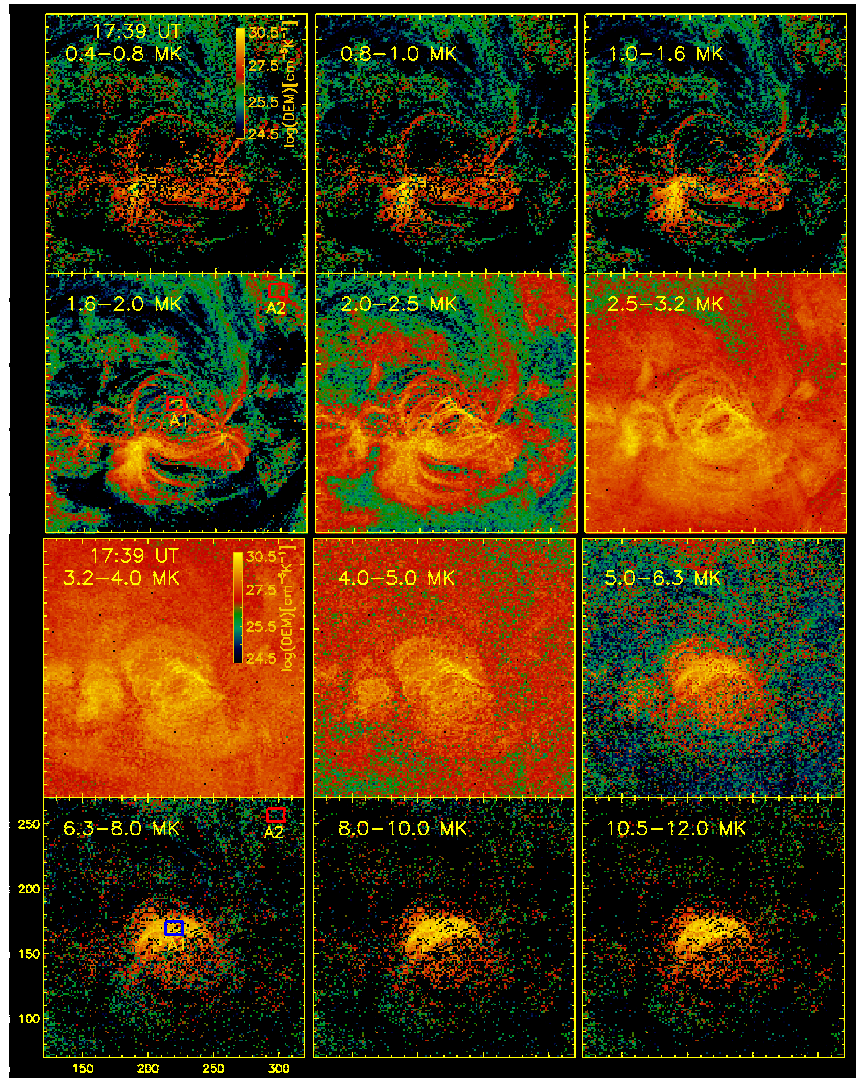

*Figure 4.* The DEM maps in the broad range of temperature for 2015 May 23 at 17:39 UT, where the *A1* and the *A2* regions are indicated by the colored boxes.

# 3. FREQUENCY DISTRIBUTION OF OSCILLATIONS IN THE FLARING REGION

Solar oscillations can be investigated with many different methods. However, the essential information about the power and dominating frequencies (or periods) is obtained by calculation of a power spectrum (see Section 2.)

All power maps marked with a letter 'A' were calculated for data, which start three hours (3h) before the flare event, and for each measured spectral line, while power maps 'B', are calculated for the data with the occurring flare. The series contained 1.5 hours of measurement. In all maps, we imposed the dark green contour, which represents the flare intensity flux at the moment of its peak, at 17:39 UT (based on GOES), and the black contour represents pores and post-flare loop.

Results are presented in the form of selected frequency bands.

In the Figures 5, 6 (A, B) we present spectral lines Ni I 5435 Å and Fe I 6302 Å, formed at heights between 200 and 300 km in the solar atmosphere.

In the Figure 5. (A, B), we present the power calculated for the line Ni I 5435 Å, which core forms around 230 km above the photosphere (Degenhardt & Wiehr (1994)), the spectral density energy distribution before and after the flare at all frequencies is very similar of the order of Log(P)=1.0 to 3.0 — suggesting that on this particular level only a small part of the energy is released into the higher parts of the atmosphere.

Comparing power maps calculated for the line Fe I 6173 Å (Fig.7), for which the core formation height lies around 276 km (Wisniewska et al. 2016), weakening of the power of oscillations after the flare can be observed, especially at the frequencies 3-4, 6-7, 8-9 and 12-14 mHz. It can also be seen that after the event, the oscillations are concentrated around the pores. The waves travel upward, guided by the magnetic field lines.

Furthermore, for the Fe I 6302 Å line, formed at 334 km (Schlaicher private communication) before the flare (see Fig. 6) energy at all frequencies accumulates in the pores. After the eruption, a part of the energy from this layer is released. As a result, the power on the map drops from Log(P)=3.8 to around 3.0.

Note at the level of around 600 km where the magnetic network is very well visible in the core of the Mg I $b_2$ line, we noticed a power debility after the flare only for the bands 10-12 and 12-14 mHz, while the power distribution for other frequencies has a random character before and after the event as well.

For this particular line, we observe the weakening of power oscillations in the foot-points (in one of the pores) of the flare.

Interesting is that on the level around 900 km in the line Na I $D_2$, the power weakening is strongly noticeable for the band 3-4 mHz in the foot-points of the flare. This effect does not occur at other frequencies.

Based on our calculations, the spectral line H$\alpha$ 6563 Å forms at about 2000 km above the photosphere. The maps labels A, calculated for this region before the event, clearly shows the power concentration in the flare area for all frequencies. However, also part of the energy is distributed in from of the surrounding fibrils. While in the context maps labeled B, calculating form data in a prior, and after the event, the power is strongly saturated only in the flare and additional the loop area. Note that the waves with frequencies 3-4 mHz propagate inside the post-flare loop.

In all analyzed lines, most power concentrates in the pores (at the foot-points of the flare), when compared to the surrounding them area. The magnetic field line structure causes this enhancement. To demonstrate the magnetic field line configuration, we calculate the extrapolation of magnetic field from SDO-magnetogram for the flaring region NOAA 12353 (described in Section 3 of this work), and we present it in the Figure \ref{extrapolation}. Based on the extrapolation result, we see that a lot of the magnetic field lines branch out from the two measured pores. Besides, several of these lines connect to the second pore with opposite magnetic field configuration in this active region. According to these structures, the waves follow the magnetic field lines and propagate along them.

It can be seen that in the lower chromosphere, the oscillations for almost all frequency ranges are concentrated mostly in the pore. We observe the most significant power enhancement for 5-minute oscillations (3-4 mHz), which propagate in this height along the magnetic field lines and begin in the foot-points of the pores. The increase of power is observed in the range of 8-9 mHz surrounding the area of the pores, but none of these waves has any substantial contribution to the flare process (black contour).

*Figure 5.*

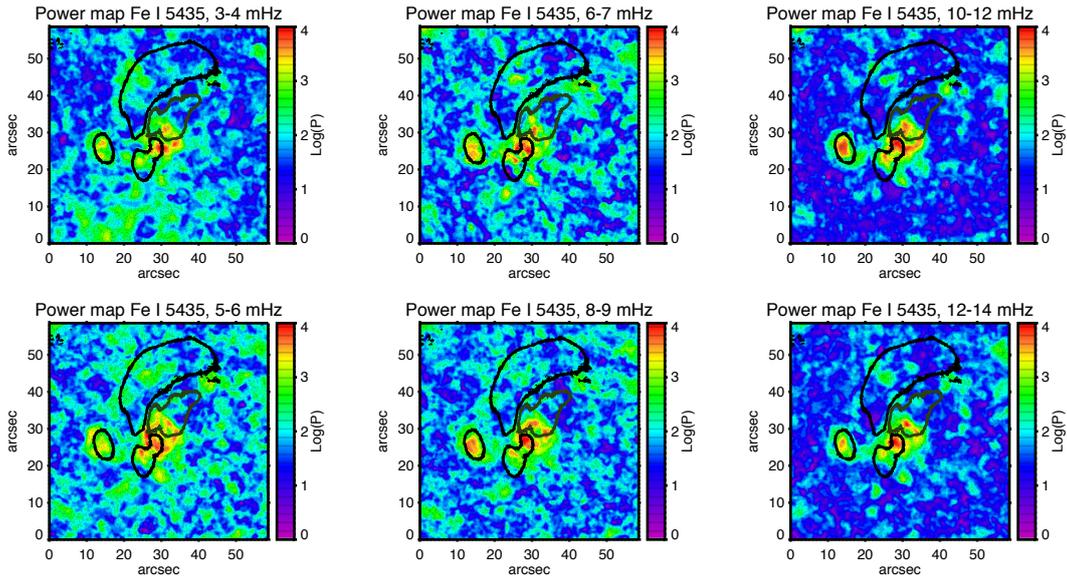

**A)** Six power maps of frequency distribution for the line **Ni I 5435 Å**, three hours before the flare.

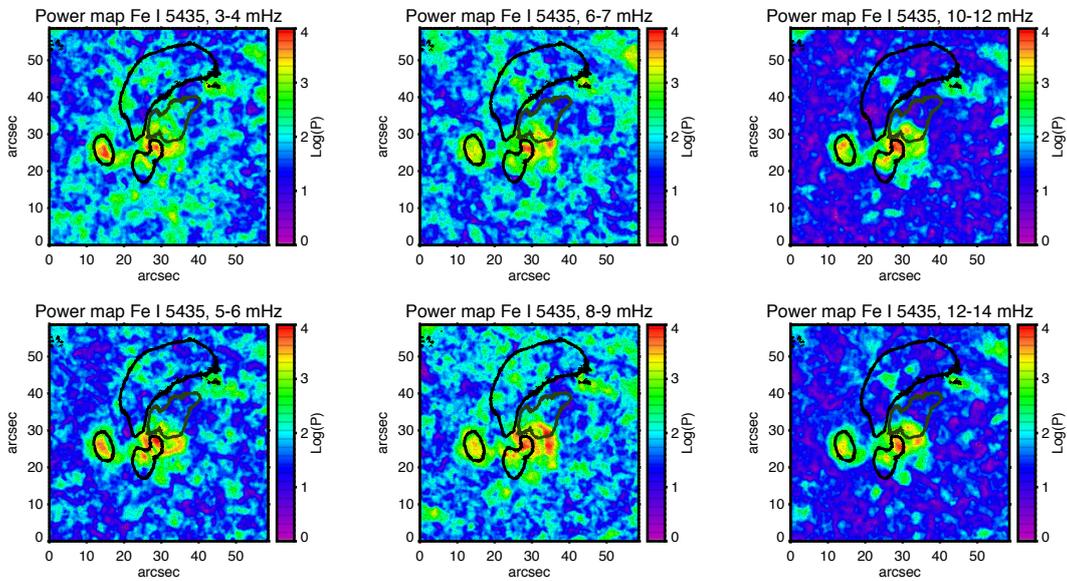

**B)** Six power maps of frequency distribution after the flare event for a chosen photospheric line formed in around 230 km above the solar surface. The power of oscillation for line **Ni I 5435 Å** is shown for selected frequency bands. The power is given by Equation 1. and calculated for each pixel of the map and plotted in logarithmic scale. The dark green contour indicates the flare intensity and position, at the peak time at 17:39 UT. The black contour marks the post-flare loop position. The atmospheric formation heights calculated for $\tau_{500}=1$, and are based on the work of Wisniewska et al.(2016).

*Figure 6.*

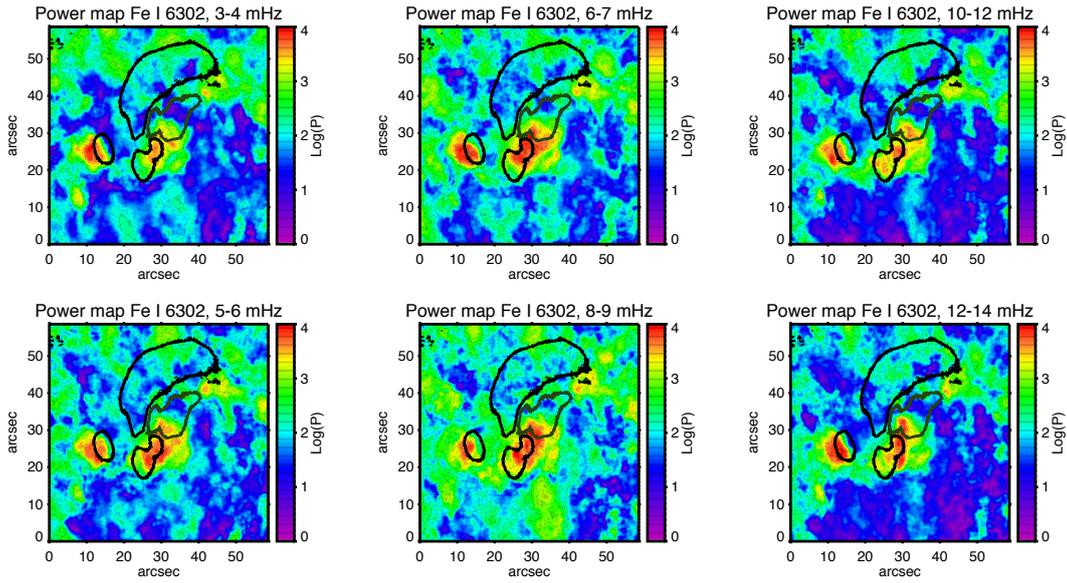

**A)** Six power maps of frequency distribution for the line **Fe I 6302 Å**, three hours before the flare.

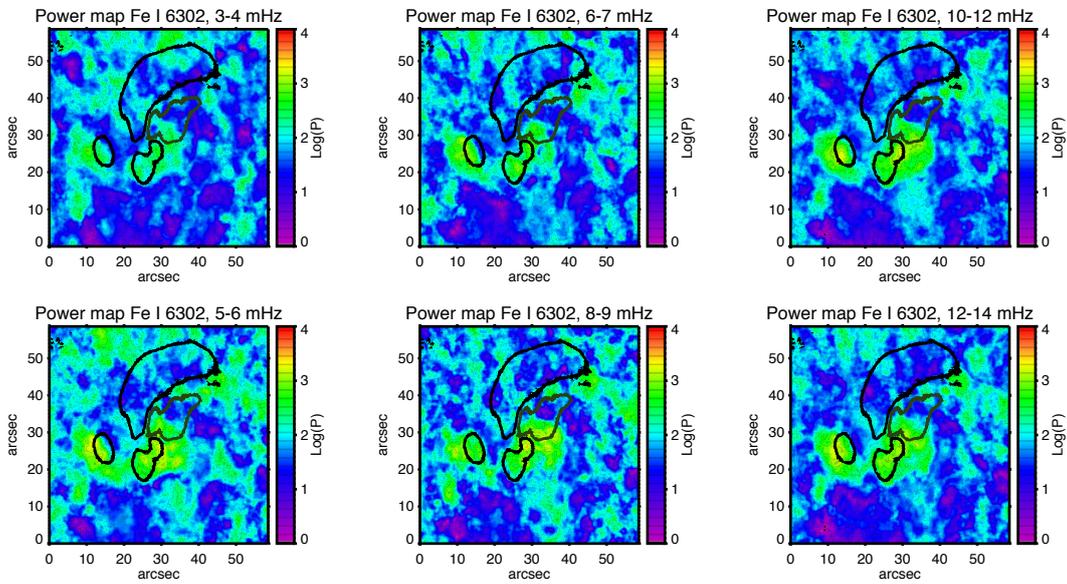

**B)** Six Power maps of frequency distribution presented after the flare for a chosen photospheric line formed in around 334 km above the solar surface. . The atmospheric formation heights calculated for $\tau_{500}=1$, and are based on the work of Wisniewska et al.(2016). The power of oscillation for line
**Fe I 6302 Å** shows selected frequency bands. The Power given by Equation 1, was calculated for each pixel of the map and plotted in logarithmic scale. The dark green contour indicates the flare intensity and position, at the peak time at 17:39 UT. The black contour marks the post-flare loop position.

*Figure 7.*

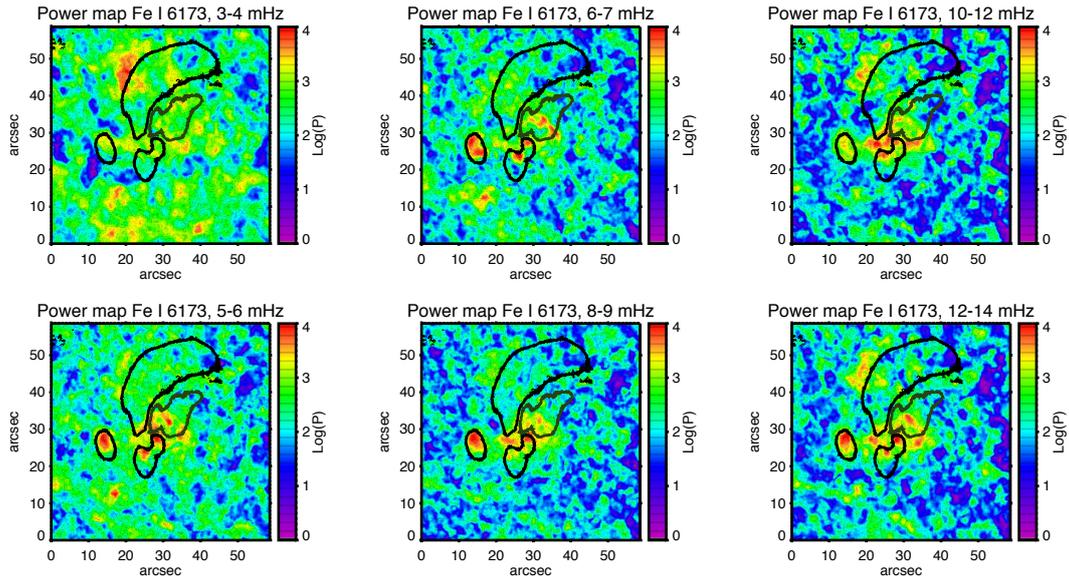

**A)** Six power maps of frequency distribution for the line **Fe I 6173 Å**, three hours before the flare.

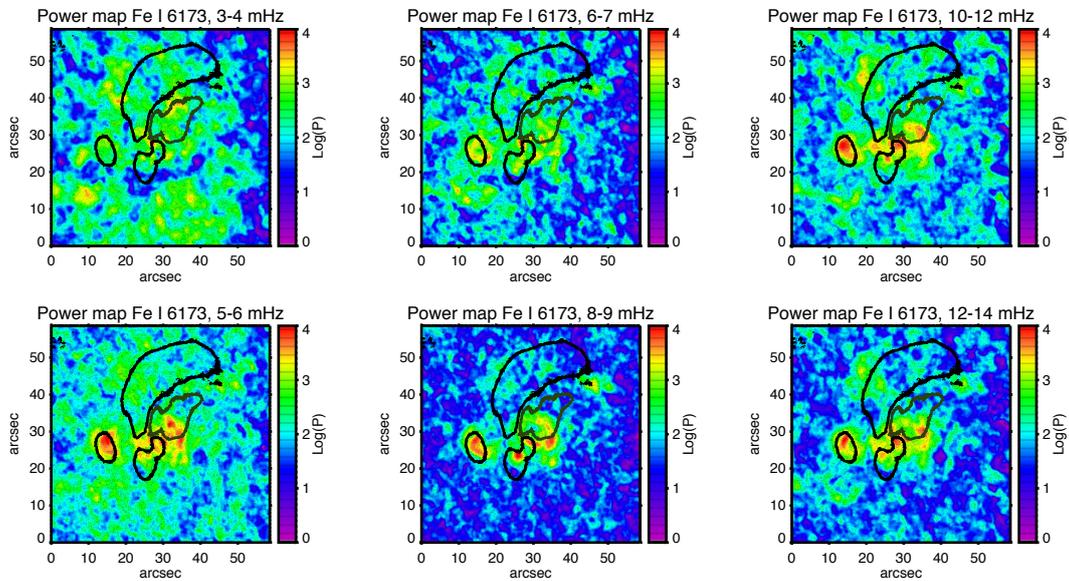

**B)** Six Power maps of frequency distribution after the flare event for a chosen photospheric line formed in around 334 km above the solar surface. The power of oscillation for line **Fe I 6173 Å** is shown for selected frequency bands. The power given by Equation 1. was calculated for each pixel of the map and plotted in logarithmic scale. The dark green contour indicates the flare intensity and position, at the peak time at 17:39 UT. The black contour marks the post-flare loop position.

*Figure 8.*

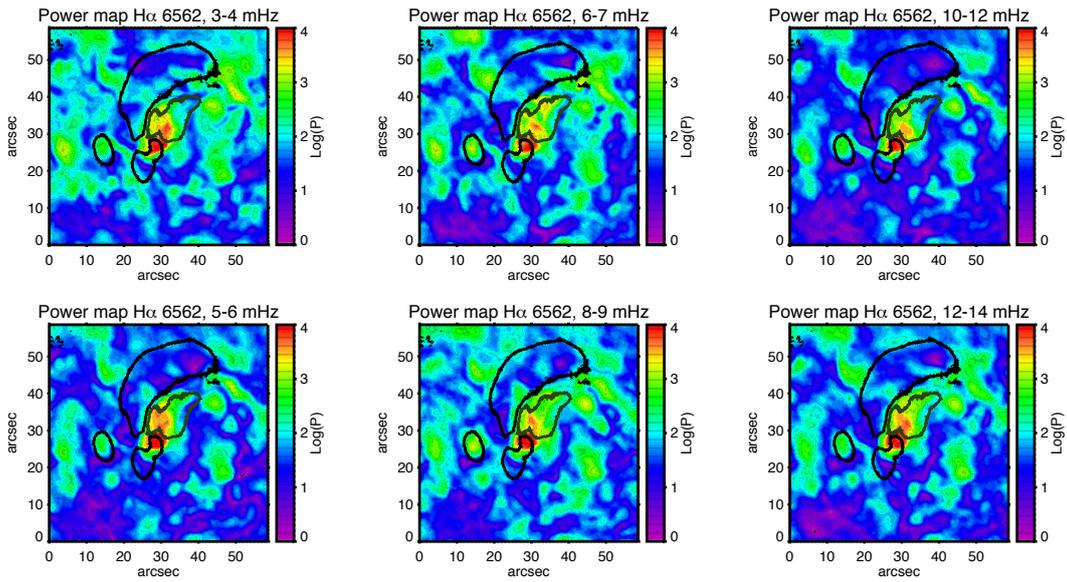

**A)** Six power maps of frequency distribution for the line **Hα 6562 Å**, three hours before the flare.

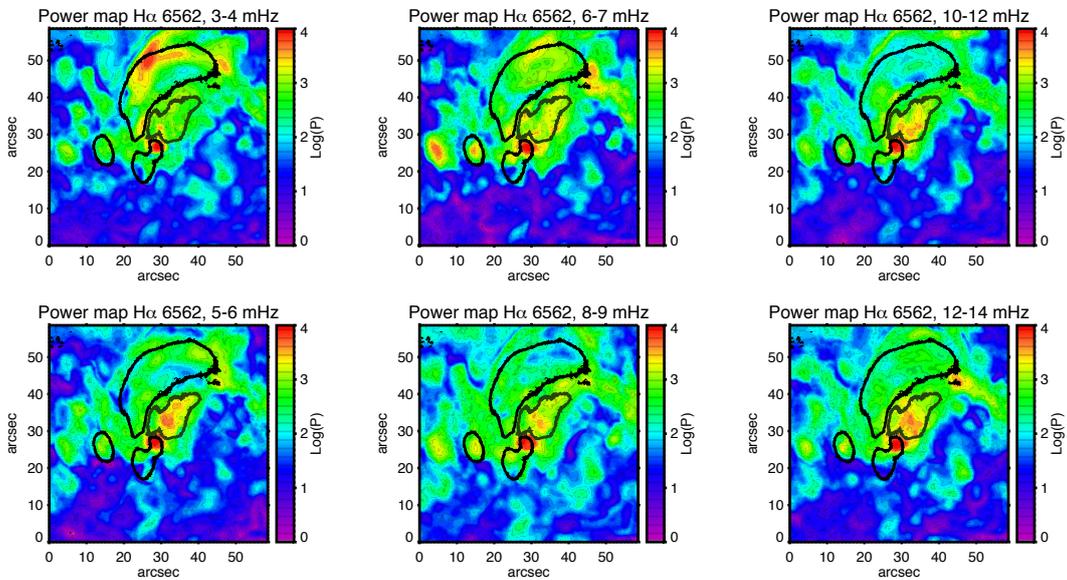

**B)** Six Power maps of frequency distribution after the flare for chromospheric line formed in around 2000 km above the solar surface. Power of oscillation for range **Hα 6562 Å** is shown for selected frequency bands. The Power given by Equation 1. was calculated for each pixel of the map and plotted in logarithmic scale. The dark green contour indicates the flare intensity and position, at the peak time at 17:39 UT. The black contour marks the post-flare loop position.

*3.1. Distribution of Low Frequency Oscillations*

In Monsue et al. (2016), the low-frequency range was studied (1-2 mHz). The results presented by the authors, for the intensity images in Hα line, show the temporal evolution of chromospheric oscillations in the flaring region for frequencies from 0 to 8.0 mHz. The temporal cadence for the data of that work involved three different sub-regions with occurring X- and M-class flares. They found, that power at the frequency band from 1.0 to 2.0 mHz is substantially enhanced in the core of the flare at the flare moment and shortly after the event. Another finding was that the power at all frequencies up to 8.0 mHz is depleted at flare maximum. This depletion is both frequency- and time-dependent. Such power variations doesn't occur in the outer surrounding area.

In the case of the C 2.3 class flare, measured by HELLRIDE, the power distribution of oscillations for the Hα confirms the finding for low-frequency waves presented in Monsue et al.2016.

Figure 9 demonstrates the low-frequency oscillations in two selected frequency bands: 0-1.0 and 1-2.0 mHz. These power maps were calculated three hours before the flare was occurring. The waves with frequencies between 0-1.0 mHz propagate inside the pore area and in the base of the flare (see left panel of Fig.9).

For the frequencies of 1.0-2.0 mHz the power enhancement for these oscillations in the core is comparable to that in the bands (3.0-14.0 mHz) discussed at the beginning of this section.

In Figure 9, we present the power maps calculated at the flare moment (at 17:39 UT) and shortly after the flare. The periodograms show the core of the flare and the area of the post-flare loop, and enhancement of the power oscillation, in both ranges from 0 to 2.0 mHz. However, the dominant frequencies of the propagating waves lie between 1.0-2.0 mHz. This result is entirely in agreement with the findings obtained by Monsue et al. (2016). The explanation for this effect is that the acoustic energy is being converted into thermal energy at flare maximum.

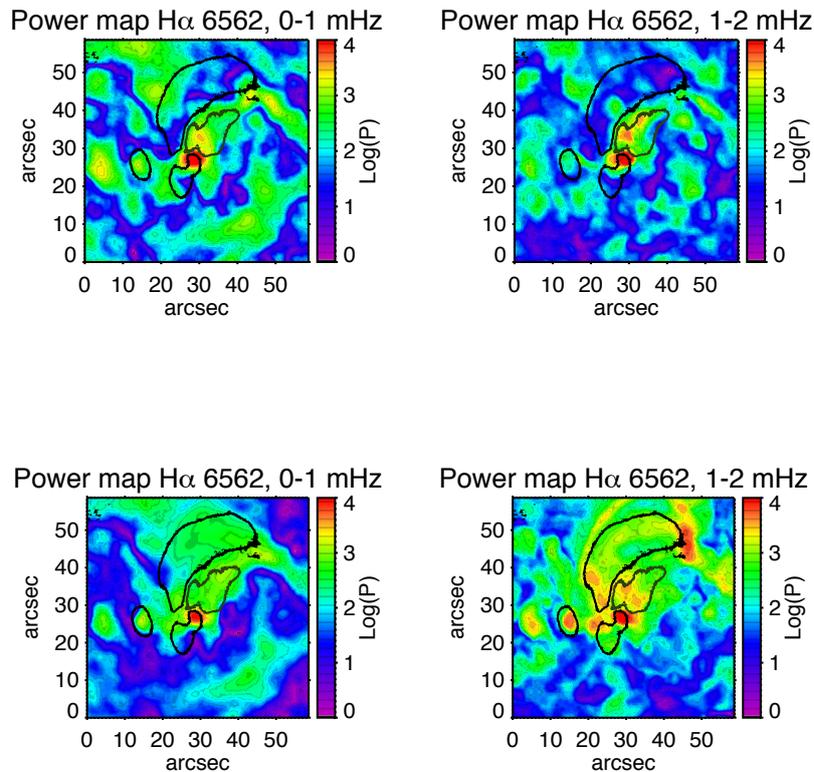

*Figure 9.* The power maps of frequency distribution obtained for line Hα Å, due (at maximum at 17:39 UT)(left) and after (right)the event. The maps are divided into two frequency bands, right: 0-1.0 mHz, and left: 1-2 mHz. The dark green contour indicates the flare intensity and position, at the peak time at 17:39 UT. The black outline marks the post-flare loop position.

## 4. MAGNETIC FIELD EVOLUTION

In the past, several authors observed the magnetic field inclination changes after flare events (Hudson (2008), Wang & Liu, (2010)). Information about the magnetic field configuration can be easily obtained using magnetograms from SDO. Both the placement of the magnetic field lines around the active area and their polarity as well may have a significant impact on the occurrence and development of the flare in the level of solar corona. In the Figure 10, we present the magnetic field evolution in the active region. We see mixed (positive and negative) polarities of the magnetic field in one of the pores. As reported by Toriumi et al. (2017), more than 80% of the sunspots, in which magnetic field configuration has mixed polarity and exhibits a σ-type configuration, lead to the eruptive events. On the magnetograms of AR 12353, (see Fig.10) there is a characteristic σ-form between two opposite polarities. This rapid evolution of the magnetic field configuration was leading to the magnetic reconnection and afterward caused the C-class flare, which occurred at 17:39 UT. Based on the data from SDO, the energy released during this C 2.3 class flare is estimating as 1.6 erg s$^{-1}$.

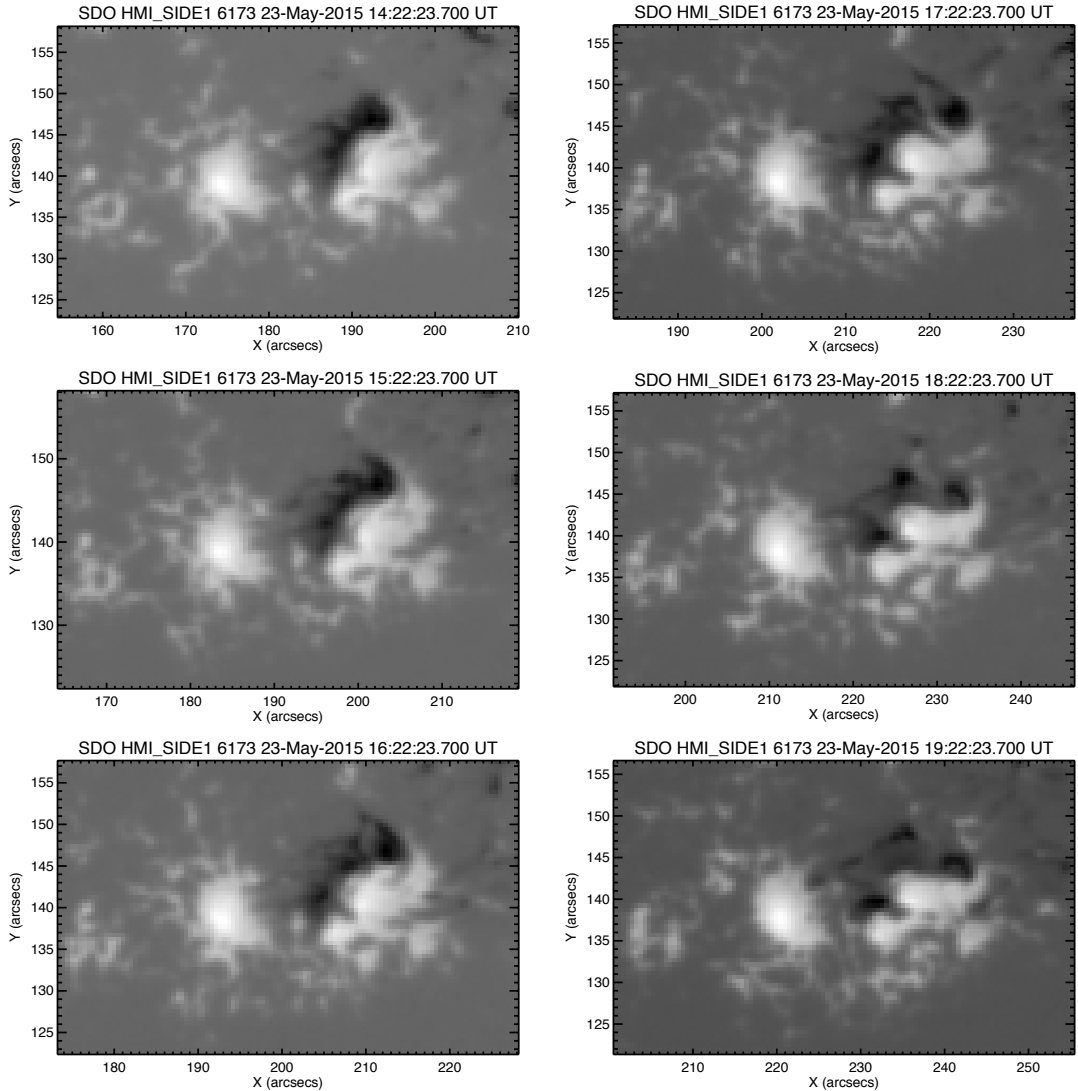

*Figure 10.* SDO/HMI HARP map of magnetic field evolution presented as magnetograms of an active region NOAA 12353. The time interval between the magnetograms is about one hour at the beginning of the measurement, which starts at 14:36 UT. The flare maximum occurs at 17:39, based on the information taken from GOES satellite. Black colour indicates a negative magnetic field, and white mark a positive value magnetic field.

## 4.1. Extrapolation of the magnetic field

This specific erupting active region requires the specification of the magnetic field at many positions (see Fig. 10 and Fig. 11).

Because the magnetic field lines of the flare loo, have their roots in a flare ribbon, thus can be disturbed due to the phenomenon. Information from magnetograms is needed to analyze the wave propagation, before and shortly after the flare. Determining at which height and area of the active region the waves start to release their energy to the higher layers of the atmosphere is crutial.

Therefore, we used the force-free magnetic field extrapolation method (Fig.11) for the constant-$\alpha$, introduced in the work of Alissandrakis (1981). Most general assumptions of for the solar atmosphere are that the currents run parallel to the field (free field conditions).

This assumption implies that:

$$\vec{\nabla} \times \vec{B} = \alpha \vec{B} \qquad (3)$$

Ignoring the displacement of the current, and in the general case, $\alpha$ is a function of position, and $B$ is the magnetic field vector.

The solution is calculated using a Fast Fourier Transform (FFT), which is a high-speed and efficient type of computation. Following Alissandrakis (1981), the $z$-axis is defined in the way to be perpendicular to the solar surface, while the boundary conditions are specified for the $z=0$ plane. The solution is given only for the upper half-space where $z > 0$. The algorithm considers the Fourier transform for each component of the magnetic field vector $x,y$ and the height $z$:

$$\hat{B}_{x(u,v,z)}, \hat{B}_{y(u,v,z)}, \hat{B}_{z(u,v,z)},$$

where the $u$ and $v$ are variables of the Fourier Transform and the B denotes a FFT. Moreover, the FFT solutions decrease with the height $z$. The algorithm creates the tables with the variables $(u,v)$ in the FT domain, and then calculates the FT components of $G$ and $B$. Next, it produces an inverse transformation of the F.T. components of the magnetic field. The complete solution of the magnetic field extrapolation is described by Alissandrakis (1981). Observational data are taken to obtain the extrapolation solutions, coming from the SDO. The HARPS time series for the active region NOAA12353. The algorithm for the magnetic field line generation uses the FLOW3 function under the SSW and is based on the three components of the magnetic field. The algorithm provided by F. Zucarello (F-Chroma Workshop, Wroclaw, Poland, 2016).

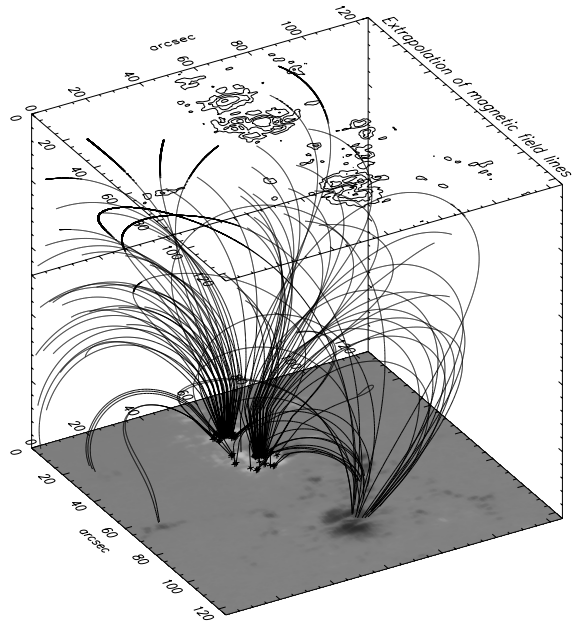

*Figure 11.* 3D-extrapolation of magnetic field, for the flaring active region in the gradual phase. The extrapolation algorithm based on the SDO/HMI magnetogram was taken at the 17:45 for the NOAA 12353. The upper contour is the contour of the magnetogram for the region.

# 5. CONCLUSIONS AND DISCUSSION

We have presented the direct observational evidence of high-frequency wave propagation during a solar flare event. The observations have been performed using HELLRIDE located at the VTT. In this work, we analyzed only the power maps measured in seven spectral regimes (covering the photospheric and chromospheric level of the atmosphere) and additionally as a context data from SDO. The SDO images contain information about the coronal plasma conditions where the flare occurs. Therefore, we performed a DEM analysis for the SDO data.

Based on the power map analysis we conclude that the high-frequency waves ($\nu >5$ mHz) in the photosphere, before and after the flare, are concentrated mainly in the foot-point of the flare loop. This area aligns with visible pores.

In the regime of the chromosphere represented by the H$\alpha$ line before the event, we observe a power enhancement, which manifests as the maximum of visible flare emission. This situation rapidly changes after the eruption and the strengthening at all frequencies is found only in the solar loop area. Based on the magnetograms and the magnetic field extrapolation calculated for this event, we have found that the post-flare-loops that arise, are guided by the magnetic field lines. The same the high-frequency waves propagate along these fine structures shortly after the phenomenon. We conclude that before the flare the power of oscillation in the pore's surrounding area has a random character, while afterward the oscillation power is concentrated mostly inside the pores, and weakens in outside of them. Moreover, the analysis of DEMs revealed that the thermal properties of in a solar loop, where high-frequency power enhancement was recognized, changed rapidly during the duration of the flare;

We identified the high frequencies in the mentioned loop, which shows the double components of DEM profiles generated for this area. The emission at high temperature (above 10 MK) was distinctly higher during the maximum of the flare phenomenon. This rapid change of the plasma temperature also influences the cutoff frequency for the waves propagating through this particular area. The change in the cutoff frequency allows waves to propagate along the post-flare loop.

Additionally, based on the 3D visualization (see Appendix A), one can conclude that before the flare, there are different dominant frequencies for the waves at characteristic atmospheric heights. Table 1, shows which frequencies of oscillation have dominant power at given heights above the photosphere before the flare.

*Table 1.* Dominating power before the flare

| **Frequency Range (mHz)** | **Height Before the Flare** | **Height After the Flare** |
|---|---|---|
| 3.0-4.0;  5.0-6.0;  6.0-7.0 | $\approx$ 300, 600, 700 | $\approx$ 300, 350, 600, 700 |
| 8.0-9.0 | $\approx$ 300, 350, 600 | $\approx$ 350, 600 |
| 10.0-12.0 | $\approx$ 600 | $\approx$ 300, 350, 600 |
| 12.0-14.0 | $\approx$ 600 | $\approx$ 600 |

Shortly after the phenomenon (up to 1.5 hr after the flare) we observe a shift of the dominant power, especially at 10-12 mHz, towards lower layers of the solar atmosphere. Right column in Table 1 shows which frequencies of oscillation have dominant power at given heights above the photosphere after the flare event.

Moreover, for all frequencies we observe a lot of power at the height of 2000 km where the loop is forming.

Concluding, the region form 300 km to roughly 600 km in the solar atmosphere is the region where the propagation of all frequencies is the strongest. This effect might be a consequence of the presence of the cut-off frequency for the waves, which is directly related to the plasma density. After the strong eruption, the plasma conditions are affected, and as a consequence the cutoff frequency for the waves can change. The analysis shows that after the flare, in the photospheric region (from 300 km to roughly 600 km) the properties of the plasma allow very high frequencies of 10-12 mHz to form powerful standing waves. In the higher region (900-2000 km) of the solar atmosphere, the power enhancement for waves occurs only along the magnetic field lines, where plasma is frozen into the field.

In summary, the understanding the frequency distribution of propagating oscillations due to a flare requires more comparable observations. Also, the numerical simulations in this field could show new ways of interpreting of the results. High-frequency waves start to be visible on the periodograms from about two hours before the flare event, in a place of its occurrence. This, effect can be easily identified with high precision, using simple FFT. Thus the prediction of the solar events like flares and the CME's may be based only on the seismic maps of the solar disk.

*Special thanks are dedicated to J.Staiger for scientific and manual support of the observations. Moreover, the authors would like to thank the F.Zucarello, and H.Schleicher for useful discussions and providing programs and additional calculations. The authors thank the referees for comments on the manuscript. The research leading to the presented results has received funding from the European Research Council under the European*
*Union's Seventh Framework Program (FP/2007-2013)/ERC grant Agreement No. 307117.*
*M.R. and A.W. acknowledge support from the SOLARNET project, which has received funding by the European Commissions FP7 Capacities Programm for the period 2013 April-2017 March, under the grant Agreement No. 312495.*

*APPENDIX*
*A. Appendix Information*
Figure 12 and 13 in this additional appendix are bases on the results presented in Section 3. The 3D visualization helps to illustrate the propagation of the acoustic waves through the atmospheric layers. The plots are divided into six different frequency ranges. Thanks to this operation, it is possible to trace particular frequency from the bottom of the photosphere up to the chromospheric level.

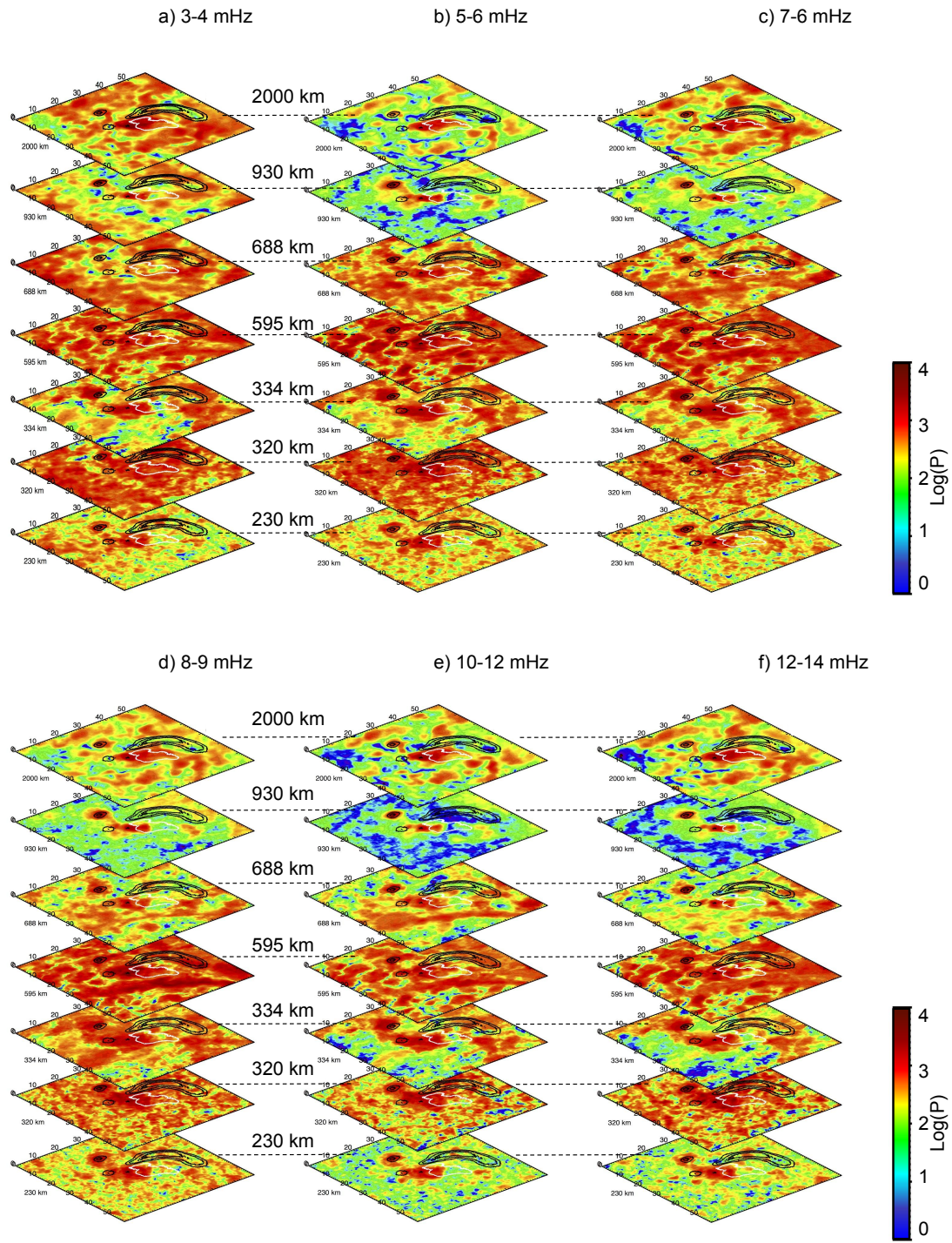

*Figure 12.* Power maps of frequency distribution before the flare. The power of oscillations is presented for selected frequency bands: a) 3-4 mHz, b) 5-6 mHz, c) 7-6mHz, d) 8-9 mHz, e) 10-12 mHz and f) 12-14 mHz. Each panel contains information from spectral lines formed at the 230 km ( Ni I 5435 Å), 320 km (Fe I 6173 Å), 334 km ( Fe I 6302 Å), 595 km (Mg I b$_2$ 5172 Å), 688 km (Fe I 5434 Å), 930 km (Na I D$_2$ 5890 Å), 2000 km (Hα 6562 Å).

The white contour indicates the flare intensity and position, at the peak time at 17:39 UT. The black contour marks post-flare loop during the gradual phase of the event. The atmospheric formation heights were calculated for τ$_{500}$=1, and are based on the work of Wisniewska et al.(2016).

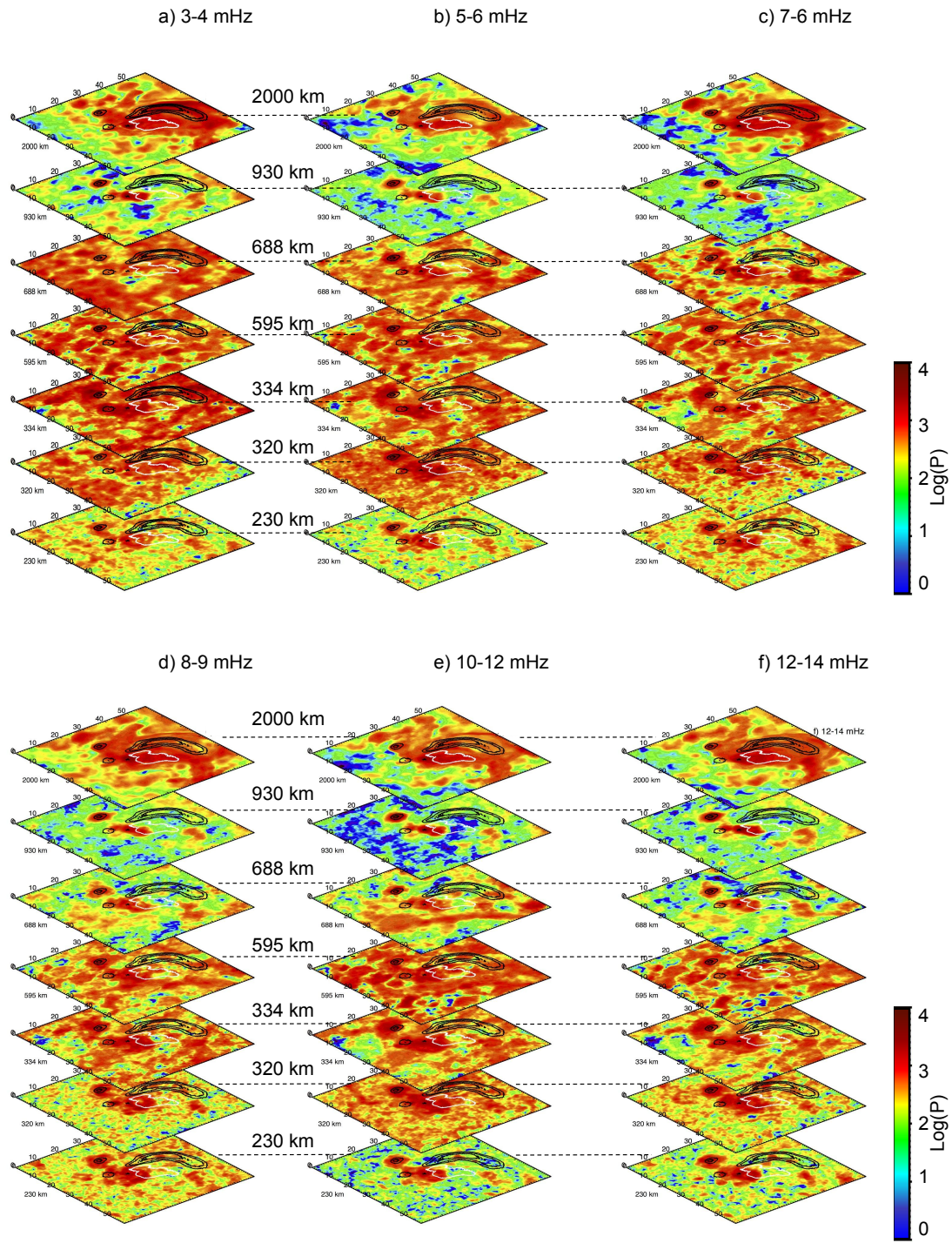

*Figure 13.* Power map of frequency distribution in prior and shortly after the flare. The power of oscillations is presented for selected frequency bands: a) 3-4 mHz, b) 5-6 mHz, c) 7-6mHz, d) 8-9 mHz, e) 10-12 mHz and f) 12-14 mHz. Each panel contains information from spectral lines formed at the 230 km (Ni I 5435 Å), 320 km (Fe I 6173 Å), 334 km ( Fe I 6302 Å), 595 km (Mg I b$_2$ 5172 Å), 688 km (Fe I 5434 Å), 930 km (Na I D$_2$ 5890 Å), 2000 km (Hα 6562 Å).
The white contour indicates the flare intensity and position, at the peak time at 17:39 UT. The black contour marks post-flare loop during the gradual phase of the event. The atmospheric formation heights were calculated for τ$_{500}$=1, and are based on the work of (Wisniewska et al.2016).